\documentclass{aastex631}
\usepackage{subfigure}
\usepackage{graphicx}
\usepackage{multirow}
\usepackage{epstopdf}
\usepackage{float}
\usepackage[normalem]{ulem}
\usepackage{array}
\usepackage{threeparttable}
\usepackage{footnote}
\usepackage[figuresright]{rotating}
\useunder{\uline}{\ul}{}

\newcommand{\qfe}{$Q_{Fe}$}
\newcommand{\velunit}{km~s$^{-1}$}

\newcommand{\ahe}{$A_{He}$}

\shorttitle{The contribution and FIP bias of three types of ICME materials with different flare intensities}
\shortauthors{Fu et al.}

\graphicspath{{./}{figures/}}

\begin{document}

\title{The contribution and FIP bias of three types of materials inside ICMEs associated with different flare intensities}

\correspondingauthor{Hui Fu}
\email{fuhui@sdu.edu.cn}

\author[0000-0002-8827-9311]{Hui Fu}
\affiliation{Shandong Key Laboratory of Optical Astronomy and Solar-Terrestrial Environment, Institute of Space Sciences, Shandong University, Weihai, Shandong, 264209, China}

\author{Xinzheng Shi}
\affiliation{Shandong Key Laboratory of Optical Astronomy and Solar-Terrestrial Environment, Institute of Space Sciences, Shandong University, Weihai, Shandong, 264209, China}

\author[0000-0002-2358-5377]{Zhenghua Huang}
\affiliation{Shandong Key Laboratory of Optical Astronomy and Solar-Terrestrial Environment, Institute of Space Sciences, Shandong University, Weihai, Shandong, 264209, China}

\author{Youqian Qi}
\affiliation{Shandong Key Laboratory of Optical Astronomy and Solar-Terrestrial Environment, Institute of Space Sciences, Shandong University, Weihai, Shandong, 264209, China}

\author[0000-0001-8938-1038]{Lidong Xia}
\affiliation{Shandong Key Laboratory of Optical Astronomy and Solar-Terrestrial Environment, Institute of Space Sciences, Shandong University, Weihai, Shandong, 264209, China}

%
%
%

\begin{abstract}
The relationship between CMEs and flares is an important issue in solar and stellar physics. The studies on the origination and generation mechanisms of ICME materials are crucial for understanding the connection between CMEs and flares.
The materials inside ICMEs can be classified into three types, coming from corona directly (corona-materials), heated by magnetic reconnection in corona (heated-corona-materials), and generated by chromospheric evaporation (chromospheric-evaporation-materials).
Here, the contribution and First Ionization Potential (FIP) bias of three types of materials inside ICMEs associated with different flare intensities are analyzed and compared.
We find that the speeds and scales of near-Earth ICMEs both increase with flare intensities.
The proportions of heated-corona-materials are nearly constant with flare intensities. The contributions of corona-materials (chromospheric-evaporation-materials) are significantly decreased (increased) with flare intensities. More than two-thirds of materials are chromospheric-evaporation-materials for ICMEs associated with strong flares.
The FIP bias of corona-materials and heated-corona-materials is almost the same. The FIP bias of chromospheric-evaporation-materials is significantly higher than that of corona-materials and heated-corona-materials, and it is increased with flare intensities.
The above characteristics of FIP bias can be explained reasonably by the origination and generation mechanisms of three types of ICME materials.
The present study demonstrates that the origination and generation mechanisms of ICME materials are significantly influenced by flare intensities.
The reasons for the elevation of FIP bias, if ICMEs are regarded as a whole, are that the FIP bias of chromospheric-evaporation-materials is much higher, and the chromospheric-evaporation-materials contributed significantly to the ICMEs which associated with strong flares.

\end{abstract}


\section{Introduction}\label{sec:intro}
The relationship between coronal mass ejections (CMEs) and flares is an important issue in solar physics and space science
\citep{1976SoPh...48..389G, 2001ApJ...559..452Z, 2003AdSpR..32.2425H, 2004SoPh..222..115L}.
The CMEs and flares are also fundamental research objects in space weather studies as they are the
most violent energy-release processes on the Sun \citep{2012LRSP....9....3W}.
It is believed that the flares and CMEs are closely connected with each other. However, they are not corresponding one by one, and they are not regarded as one being the cause and the other as the effect
\citep{1995A&A...304..585H, 2012LRSP....9....3W}.
The statistical analysis demonstrates that the higher the flare class,
the higher the CME association rate \citep{2006ApJ...650L.143Y, 2007ApJ...665.1428W}.
There are about 10\% (40\%) of C-class (M-class) flares are associated with CMEs, and more than 80\% of X-class flares are accompanied by CMEs \citep{2006ApJ...650L.143Y, 2007ApJ...665.1428W}.
The studies on the connection between flares and CMEs are crucial for understanding the nature of the two phenomena \citep{1983SoPh...89...89M, 1986A&A...162..283H, 2001ApJ...559..452Z, 2012LRSP....9....3W, 2022ApJ...940..103S}.

The speeds and accelerations of CMEs associated with and without flares are significantly different. Generally, the fastest and most energetic kinds of CMEs are associated with strong flares \citep{2002ESASP.508..409W, 2007JGRA..112.6112G}.
Statistically, the interplanetary coronal mass ejections (ICMEs) accompanied by flares are faster than those not associated with flares on the Sun \citep{1976SoPh...48..389G, 1999JGR...10424739S, 2002ApJ...581..694M, 2006A&A...456.1153C, 2017ApJS..229...29W, 2019SSRv..215...39L, 2022ApJ...940..103S}.
The reason for the ICMEs associated with flares are fast is that the CMEs can be accelerated extra by the magnetic reconnection during flares \citep{2007SoPh..241...99M, 2015ApJ...804L..38S, 2018ApJ...857L..21S, 2020ApJ...893..141Z}.
Based on observations by the LASCO coronagraph onboard SoHO, the CMEs over Solar Cycles 23 and 24 are analyzed \citep{2019SSRv..215...39L}. The statistical results show that the CMEs associated with flares have larger kinetic energy, angular width, mass, and higher speed. \citet{2017ApJS..229...29W} reconstruct the 3D structure of CMEs based on observations taken by STEREO and SoHO coronagraphs. Hence, the acceleration and speed of CMEs can be derived more credibly. The results demonstrate that the CMEs associated with flares are faster than those not accompanied by flares. This characteristic is more clear below 10 R$_{\bigodot}$ (see figure 4. (b) in \citeauthor{2017ApJS..229...29W}, \citeyear{2017ApJS..229...29W}). 

Statistically, the charge states, helium abundance, and FIP bias are all higher than those inside the background solar wind \citep{2003JGRA..108.1156C, 2004JGRA..109.9104R, 2008ApJ...682.1289R, 2010SoPh..264..189R, 2017SSRv..212.1159M, 2018SoPh..293..122O, 2021SoPh..296..111S, 2022ApJ...940..103S}.
The charge states inside ICMEs associated with flares are higher than those of ICMEs not accompanied by flares \citep{2001JGR...10629231L, 2004JGRA..109.1112L, 2017SSRv..212.1159M, 2022ApJ...940..103S}.
In addition, the statistical results show that the charge states inside ICMEs are positively correlated with flare intensities \citep{2013SoPh..284...17G, 2022ApJ...928..136Z}.
The above results provide further evidence that the materials inside ICMEs are strongly heated during flaring processes.
Early studies indicate that the higher helium abundance inside ICMEs has a close relationship with large solar flares \citep{1972SoPh...23..467H, 1982JGR....87.7370B}.
The statistical analysis demonstrates that the helium abundance inside ICMEs is also positively correlated with flare intensities \citep{2022ApJ...928..136Z}.

The in-situ properties of ICMEs are closely related to the origination and generation mechanisms of the materials on the Sun. The charge states and element abundance (\ahe, FIP bias) inside ICMEs carry the information near the Sun, where the materials are heated and injected into CMEs \citep{2017SSRv..212.1159M, 2020ApJ...900L..18F, 2022ApJ...928..136Z}.
There is a consensus that the materials associated with higher charge states inside ICMEs are heated during the flaring process \citep{2001JGR...10629231L, 2002A&ARv..10..313P, 2004JGRA..109.1112L, 2006SSRv..124..145G, 2013SoPh..284...17G, 2005ApJ...622.1251L, 2013NatPh...9..489S, 2015ApJ...808L..15S, 2016ApJS..224...27S}.
The reason for higher \ahe\ in some of ICMEs is not clearly known \citep{2017SSRv..212.1159M}.
\citet{1997GMS....99..245N} suggest that the lower solar atmosphere materials with higher \ahe\ can be transported into CMEs by an unknown mechanism.
The reason for suggesting the higher \ahe\ materials coming from the lower atmosphere is that the \ahe\ is depleted in the corona \citep{2001ApJ...546..552L, 2003ApJ...591.1257L}.
In addition, the \ahe\ inside solar wind is also lower than that in the photosphere \citep{2001GeoRL..28.2767A, 2007ApJ...660..901K, 2012ApJ...745..162K, 2018MNRAS.478.1884F}.
{Recently, the origination and generation mechanisms of ICME materials have been analyzed in a series of studies implemented by \citet{2020ApJ...900L..18F}, \citet{2022ApJ...928..136Z}, and \citet{2022ApJ...940..103S} in the same group.
Firstly, \citet{2020ApJ...900L..18F} find that the materials heated by the chromospheric evaporation process at flux rope footpoint regions can be injected into CMEs.
The above case study demonstrates that the hot plasma inside ICMEs is not only heated by magnetic reconnection between the flare loops and flux ropes, but also can be produced by the chromospheric evaporation process at flux rope footpoint regions during the flaring processes.
The origination and generation mechanisms of the materials inside ICMEs can be deduced by the in-situ properties of ICMEs.
Qualitatively, the plasma associated with lower charge states and \ahe\ inside ICMEs should come from the corona directly, as the plasma is not associated with additional heating, and the helium abundance of the corona is lower \citep{2001ApJ...546..552L, 2003ApJ...591.1257L}.
The materials with higher \qfe\ and lower \ahe\ should be heated by magnetic reconnection between the outward-moving flux rope and flare loops.
In contrast, the materials associated with higher \qfe\ and higher \ahe\ should be produced by the chromospheric evaporation process at the flux rope footpoint regions as the \ahe\ is higher in the lower solar atmosphere \citep{2020ApJ...900L..18F, 2022ApJ...928..136Z}.
Therefore, The higher charge states and \ahe\ inside ICMEs can be explained reasonably by the fact that the materials can be produced by the chromospheric evaporation process at the chromosphere.}

{Secondly, the contributions of the two types of hot materials inside ICMEs are evaluated by \citet{2022ApJ...928..136Z}.
They link the ICMEs detected near the Earth and their associated activities on the Sun by observations taken by coronagraphs onboard STEREO-A and -B.
The near-Earth ICMEs and activities on the Sun can be credibly associated as the ICMEs are traced from the Sun to the Earth continuously by imaging and coronagraph observations.
However, the above criteria can only be satisfied from 2009 to 2013, when the STEREO-A and -B are suitable for tracing the ICMEs from the Sun to the Earth \citep{2017ApJS..229...29W, 2022ApJ...928..136Z}.
In addition, only the ICMEs with clear structures can be followed all the way from the Sun to the Earth. Hence, the analyzed number of ICMEs is less, and the interval is limited to 2009 to 2013 in \citet{2022ApJ...928..136Z}.
The ICMEs are then classified into two types, with flares (flare-CMEs/FCs) and without flares (Non-flare-CMEs/NFCs) on the Sun.
They find that the \qfe\ of NFCs is less than 12 and it presents a one-peak distribution with the peak locating at about QFe=10.
In contrast, the \qfe\ of FCs is significantly higher and it demonstrates a two-peak distribution with the minimum locating at about QFe=12 between the two peaks (see Figure~1 in \citeauthor{2022ApJ...928..136Z}, \citeyear{2022ApJ...928..136Z}).
The above results mean that the ICME materials associated with \qfe\ higher than 12 should be related to the flaring processes.
In addition, the distribution characteristics of FCs and NFCs in the space of \qfe\ and \ahe\ are analyzed and compared.
The NFCs are mainly located at the lower \qfe\ and \ahe\ quadrant.
This is consistent with the notion that the materials inside NFCs come from the corona directly \citep{1996ApJ...470..629H, 2000A&A...358.1097H, 2003A&A...400.1071H, 2018ApJ...863..169D}.
In contrast, the plasma origination and generation mechanisms of FCs are more complicated.
There are about 28\% of materials inside FCs come from the corona directly (with lower \qfe\ and \ahe).
The proportion of the plasma heated by magnetic reconnection in the corona (with higher \qfe\ and lower \ahe) is 19\%.
The proportion of the materials generated by the chromospheric evaporation at flux rope footpoint regions (with higher \qfe\ and higher \ahe) can reach up to 40\% in their database (see Figure~2 in \citeauthor{2022ApJ...928..136Z}, \citeyear{2022ApJ...928..136Z} for details).}

{Thirdly, the occurrence rates and properties of FCs and NFCs from 1999 to 2020 are statistically analyzed and compared by \citet{2022ApJ...940..103S}.
The ICMEs in the well-known Richardson and Cane (RC) list\footnote{\url{https://izw1.caltech.edu/ACE/ASC/DATA/level3/icmetable2.htm/}} are classified into FCs and NFCs based on the in-situ detected characteristics of \qfe.
The statistical analysis implemented by \citet{2022ApJ...928..136Z} demonstrates that the \qfe\ is less than 12 for ICMEs without flares on the Sun.
The ICME materials with \qfe\ greater than 12 should be heated during the flaring processes.
Therefore, the ICMEs whose \qfe\ for all samples is lower than 12 are categorized into NFCs.
An ICME is classified into FC if \qfe\ for more than 30\% of samples is higher than 12.
The threshold of 30\% is adopted, as the authors intend to make the classification procedure of FCs more reliable and make the group of FCs more pure.
In total, about 83\% of the ICMEs in the RC list are classified into FCs and NFCs.
The statistical analysis shows that the occurrence counts of FCs during solar cycle 23 are about three times higher than those during solar cycle 24.
Whereas, the numbers of NFCs are almost the same during solar cycles 23 and 24 (see top panel of Figure~1 in \citeauthor{2022ApJ...940..103S}, \citeyear{2022ApJ...940..103S}).
This means that the occurrence rate of FCs is strongly influenced by the solar cycle activities.
The speed, charge states, helium abundance, and FIP bias for FCs are all higher than those for NFCs. \citet{2022ApJ...940..103S} suggest that the above property differences should be related to the material sources and generation mechanisms of FCs and NFCs.
}

At present, the contribution and FIP bias of different types of ICME materials that originate from different regions and/or are generated by different mechanisms are not analyzed statistically.
Whether the speeds and scales of near-Earth ICMEs change with the flare intensities?
How do the contributions of the three types of materials (from the corona directly, heated by magnetic reconnection between the outward-moving flux rope and flare loops, and generated by chromospheric evaporation process at the flux rope footpoint regions) change with the associated flare intensities?
Does the FIP bias inside three types of materials the same or not?
Why the FIP bias is higher inside ICMEs if the ICMEs are regarded as a whole?
Where the higher FIP bias plasma inside ICMEs comes from?

In the present study, the ICMEs associated with different flare intensities are analyzed and compared. The materials inside ICMEs are first classified into three types, coming from the corona directly (corona-materials), heated by magnetic reconnection in the corona (heated-corona-materials), and generated by chromospheric evaporation processes in the chromosphere (chromospheric-evaporation-materials) based on the in-situ detections of \qfe\ and \ahe\ \citep{2020ApJ...900L..18F, 2022ApJ...928..136Z}.
Then the contribution and FIP bias of the three types of materials
are statistically analyzed and compared for the ICMEs associated with different flare intensities.
We find that the speeds and scales of near-Earth ICMEs both increase with flare intensities.
The proportions of the three types of materials inside ICMEs change significantly with the flare intensities.
The FIP bias of corona-materials and heated-corona-materials is almost the same, and the FIP bias of chromospheric-evaporation-materials is the highest. The FIP bias of chromospheric-evaporation-materials is significantly increased with flare intensities.
The present study will enhance our understanding on the origination and generation mechanisms of the ICME materials and the relationship between CMEs and flares.
The present study demonstrates that the origination and generation mechanisms of materials inside ICMEs are significantly influenced by flare intensities.
Our results also clarify the question of why the FIP bias is higher inside ICMEs in a statistical manner. The reasons for the elevation of FIP bias, if ICMEs are regarded as a whole, are that the FIP bias of chromospheric-evaporation-materials is much higher, and the chromospheric-evaporation-materials contributed significantly to the ICMEs which associated with strong flares.

The present paper is organized in the following. In Section 2, the data and analysis method is described. The statistical results are presented and discussed in Section 3. Finally, the main results and conclusion are summarized in Section 4.

\section{Data and analysis method} \label{sec:style}
In the present work, the ICMEs list, which is organized by Richardson and Cane (RC list \footnote{\url{https://izw1.caltech.edu/ACE/ASC/DATA/level3/icmetable2.htm/}}), is adopted. The RC list includes the near-Earth ICMEs since Jan 1996, and it is updated frequently. The ICMEs included in the list are chosen by both criteria characteristic of ICMEs and visual inspection of the in-situ detection, as the organizers wish to produce a comprehensive near-Earth ICMEs list \citep{2003JGRA..108.1156C, 2004JGRA..109.9104R, 2010SoPh..264..189R}. The list not only includes the ICMEs associated with standard characteristics (such as magnetic clouds, low proton temperature, and low plasma beta \citep{1990JGR....9511957L, 2017SSRv..212.1159M, 2018SoPh..293..122O}) but also some cases associated with a part of the characteristics of ICMEs are included in the list.
The ICMEs listed in the RC list from 1999 to 2011, which includes the full solar cycle 23, are analyzed in the present study.
Not all ICMEs in the RC list are analyzed, as the FIP bias detection of Advanced Composition Explorer (ACE, \citeauthor{1998SSRv...86....1S}, \citeyear{1998SSRv...86....1S}) is influenced by a space weather event after August 23 2011 \citep{2016ApJ...826...10Z}.

The in-situ parameters of near-Earth ICMEs are measured by Wind \citep{1995SSRv...71....5A} and ACE \citep{1998SSRv...86....1S} spacecraft. The speed and helium abundance are measured by the Solar Wind Experiment (SWE) Faraday cup instruments \citep{1995SSRv...71...55O} onboard Wind.
The average charge states of Iron (\qfe) and FIP bias are derived from the detection of the Solar Wind Ion Composition Spectrometer (SWICS, \citeauthor{1998SSRv...86..497G}, \citeyear{1998SSRv...86..497G}) onboard ACE.
{Previous studies demonstrate that the FIP bias for different elements is not the same \citep{2015LRSP...12....2L, 2019ApJ...879..124L}.
The FIP bias is calculated from the ratio of $((Fe+Mg+Si)/O)_{ICME}$ and $((Fe+Mg+Si)/O)_{photosphere}$ (see equation 5 in \citeauthor{2000JGR...10527217V}, \citeyear{2000JGR...10527217V}) in the present study.
The density ratio of $Fe/O$, $Mg/O$, and $Si/O$ of ICMEs is detected by the SWICS onboard ACE.
The element abundance in the photosphere is obtained from Table 1 in \citet{2009ARA&A..47..481A}.
The calculated density ratio of $Fe/O$, $Mg/O$, and $Si/O$ in the photosphere is 0.065, 0.081, and 0.066, respectively.
Therefore, the derived FIP bias represents the average FIP bias of Fe, Mg, and Si inside ICME materials.}

The previous statistical results demonstrate that the charge states of ICMEs are positively correlated with flare intensities \citep{2013SoPh..284...17G, 2022ApJ...928..136Z}. Generally, the average \qfe\ is higher (lower) than 14.0 (12.5) for the ICMEs associated with X-class (equal to or less than the C-class ) of flares. The average \qfe\ ranges from 12.5 to 14.0 for the ICMEs associated with M-class of flares (see Figure 7 in \citeauthor{2013SoPh..284...17G},\citeyear{2013SoPh..284...17G} and Figure 3(a) in \citeauthor{2022ApJ...928..136Z}, \citeyear{2022ApJ...928..136Z}).
Hence, the associated flare intensities can be inferred by the in-situ \qfe\ inside ICMEs, statistically.

In the present study, the FCs are further categorized into three types, FCs associated with average \qfe\ less than 12.5 (LQFe-FCs), average \qfe\ ranges from 12.5 to 14.0 (MQFe -FCs), and average \qfe\ higher than 14.0 (HQFe-FCs).
The procedures for the present statistical study are summarized in the following.

First, an ICME is regarded as associated with a flare (FC) on the Sun if more than 30\% of its \qfe\ is higher than 12, and the ICMEs with \qfe\ all lower than 12 are regarded as not associated with flares (NFCs) \citep{2022ApJ...928..136Z, 2022ApJ...940..103S}.
{The threshold of 30\% is adopted, as the authors intend to make the classification procedure of FCs more reliable and the group of FCs more pure.
In this case, the ICMEs with less than 30\% of their \qfe\ higher than 12 are filtered out.
The ICMEs with a duration of longer than 10 hours are only considered, as we focus on the distribution characteristics inside ICMEs.
The number of ICMEs shorter than 10 hours is 18, and 52 ICMEs are filtered out by the threshold of \qfe.
In addition, there are 8 ICMEs that the detected QFe are not valid in more than half of the samples. The above ICMEs are also filtered out.
Hence, 210 of 288 ICMEs identified between 1999-Jan-01 and 2011-Aug-23 on the well-known RC list are analyzed in the present study.
}


Second, the ICMEs associated with flares are categorized into three types, LQFe-FCs (average \qfe\ less than 12.5), MQFe-FCs (average \qfe\ ranges from 12.5 to 14.0), and HQFe-FCs (average \qfe\ higher than 14.0) based on the average \qfe. Qualitatively, the higher the average \qfe\ inside ICMEs, the stronger the associated flares, although the average \qfe\ are not precisely correlated with flare intensities \citep{2013SoPh..284...17G, 2022ApJ...928..136Z}. Hence, the three types of FCs defined in the present study should correspond to different flare intensities on the Sun. 
{Statistically, the associated flare intensities should be the lowest for LQFe-FCs and highest for HQFe-FCs, with the flare intensities of MQFe-FCs lying in between.}

{Third, the materials inside ICMEs are classified into three types based on in-situ \qfe\ and \ahe\ \citep{2020ApJ...900L..18F, 2022ApJ...928..136Z}.
The ICME materials with \qfe$<=12$ and \ahe$<=7$, \qfe$>12$ and \ahe$<=7$, and \qfe$>12$ and \ahe$>7$ are categorized into corona-materials, heated-corona-materials, and chromospheric-evaporation-materials, respectively \citep{2022ApJ...928..136Z}.}

Finally, the speeds and scales of NFCs, LQFe-FCs, MQFe-FCs, and HQFe-FCs are analyzed.
The contribution and FIP bias of three types of materials inside NFCs, LQFe-FCs, MQFe-FCs, and HQFe-FCs are analyzed and compared.

\section{Results and Discussion} \label{sec:floats}

\subsection{The speeds and scales of ICMEs associated with different flare intensities}

\begin{figure}
\centering
\includegraphics[width=1.0\textwidth]{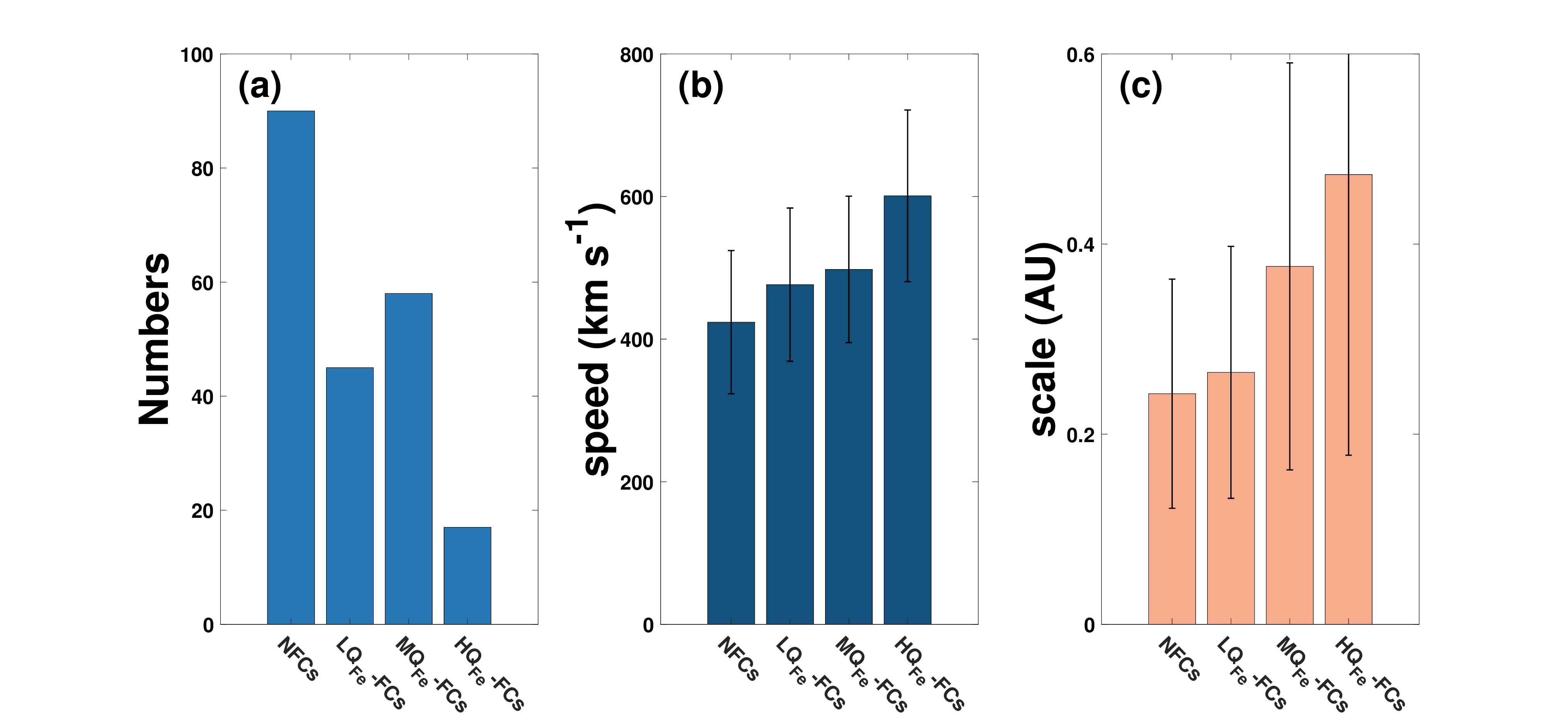}
\caption{The numbers (left panel), speeds (middle panel), and scales (right panel) of NFCs, and three types of FCs. The error bars in panels (b) and (c) represent the standard deviations of speeds and scales of ICMEs. It is clear that the speeds and scales of ICMEs are increased significantly with flare intensities.
}
\label{fig:figure1_number_speeds_scales}
\end{figure}

The numbers, speeds, and scales of NFCs and three types of FCs (LQFe-FCs, MQFe-FCs, HQFe-FCs) are presented in Figure~\ref{fig:figure1_number_speeds_scales} {and Table~\ref{tab:addlabel}}.
A total of 210 ICMEs are analyzed in the present study, and the number of NFCs is 90.
The numbers (total durations) of LQFe-FCs, MQFe-FCs, and HQFe-FCs are 45 (1377 hours), 58 (2021 hours), and 17 (589 hours), respectively {(see the second and third columns in Table~\ref{tab:addlabel})}.
The numbers and total durations of different types of ICMEs are not small, which means the relevant results should be statistically significant.

\begin{table}[htbp]
  \centering
  \caption{The numbers, speeds, and scales of NFCs and three types of FCs.}
    \begin{tabular}{ccccc}
    \hline
    \hline
          & Numbers & Total duration (hours) & Speed (\velunit) & Scale (AU) \\
    \hline
    NFCs  & 90    & 2174  & 424$\pm$101 & 0.24$\pm$0.12 \\
    L$Q_{Fe}$\--FCs & 45    & 1377  & 476$\pm$107 & 0.27$\pm$0.13 \\
    M$Q_{Fe}$\--FCs & 58    & 2021  & 498$\pm$103 & 0.38$\pm$0.21 \\
    H$Q_{Fe}$\--FCs & 17    & 589   & 601$\pm$120 & 0.47$\pm$0.30 \\
    \hline
    \end{tabular}%
  \label{tab:addlabel}%
\end{table}%

The average speeds of NFCs and three types of FCs are presented in the middle panel of Figure~\ref{fig:figure1_number_speeds_scales}.
It is clear that the average speeds of ICMEs are increased with flare intensities. Statistically, the average speeds of HQFe-FCs are significantly higher than those of NFCs, LQFe-FCs, and MQFe-FCs. The average speed of NFCs is $424\pm101$ \velunit, and the average speeds for LQFe-FCs, MQFe-FCs, and HQFe-FCs are 476$\pm$107, 498$\pm$103, and 601$\pm$120 \velunit, respectively.

The average scales of NFCs and three types of FCs is shown in the right panel of Figure~\ref{fig:figure1_number_speeds_scales}. The scales of ICMEs are significantly increased with the increase of the flare intensities. The average scale of NFCs is 0.24$\pm$0.12 AU. In contrast, the average scales of LQFe-FCs, MQFe-FCs, and HQFe-FCs are 0.27$\pm$0.13, 0.38$\pm$0.21 and 0.47$\pm$0.30 AU, respectively.

The above statistical results demonstrate that the ICMEs associated with flares (FCs) are larger and faster than those not accompanied by flares (NFCs). The fact that the FCs are faster than NFCs is consistent with the coronagraph observations. The remote observations demonstrate that the CMEs associated with flares are faster than those not accompanied by flares near the Sun \citep{1999JGR...10424739S, 2002ApJ...581..694M, 2006A&A...456.1153C, 2017ApJS..229...29W, 2019SSRv..215...39L}.
In the present study, the FCs are further categorized into three types based on the average \qfe\ inside ICMEs. Statistically, the associated flare intensities of LQFe-FCs, MQFe-FCs, and HQFe-FCs should be increased, as the previous statistical results find that the averaged \qfe\ inside ICMEs is positively correlated with flare intensities \citep{2013SoPh..284...17G, 2022ApJ...928..136Z}.
The present statistical results demonstrate that the speeds and scales of ICMEs near the Earth are both positively correlated with flare intensities, although the speeds and scales of ICMEs should be both changed during the propagation in the heliosphere.
The fact that the ICME speed is increased with flare intensities is also consistent with the notion that the CMEs are accelerated extra by magnetic reconnection during flares \citep{2007SoPh..241...99M, 2018ApJ...857L..21S, 2020ApJ...893..141Z}.
The stronger the flares, the more acceleration of CMEs, hence the faster the associated ICMEs.
On the other hand, the above statistical results also indicate that the present classification procedure of LQFe-FCs, MQFe-FCs, and HQFe-FCs is credible. The associated flare intensities of LQFe-FCs, MQFe-FCs, and HQFe-FCs should be increased statistically.

\subsection{The contributions of the three types of materials inside ICMEs that are associated with different flare intensities}

\begin{figure}[!h]
\centering
\includegraphics[width=1.0\textwidth]{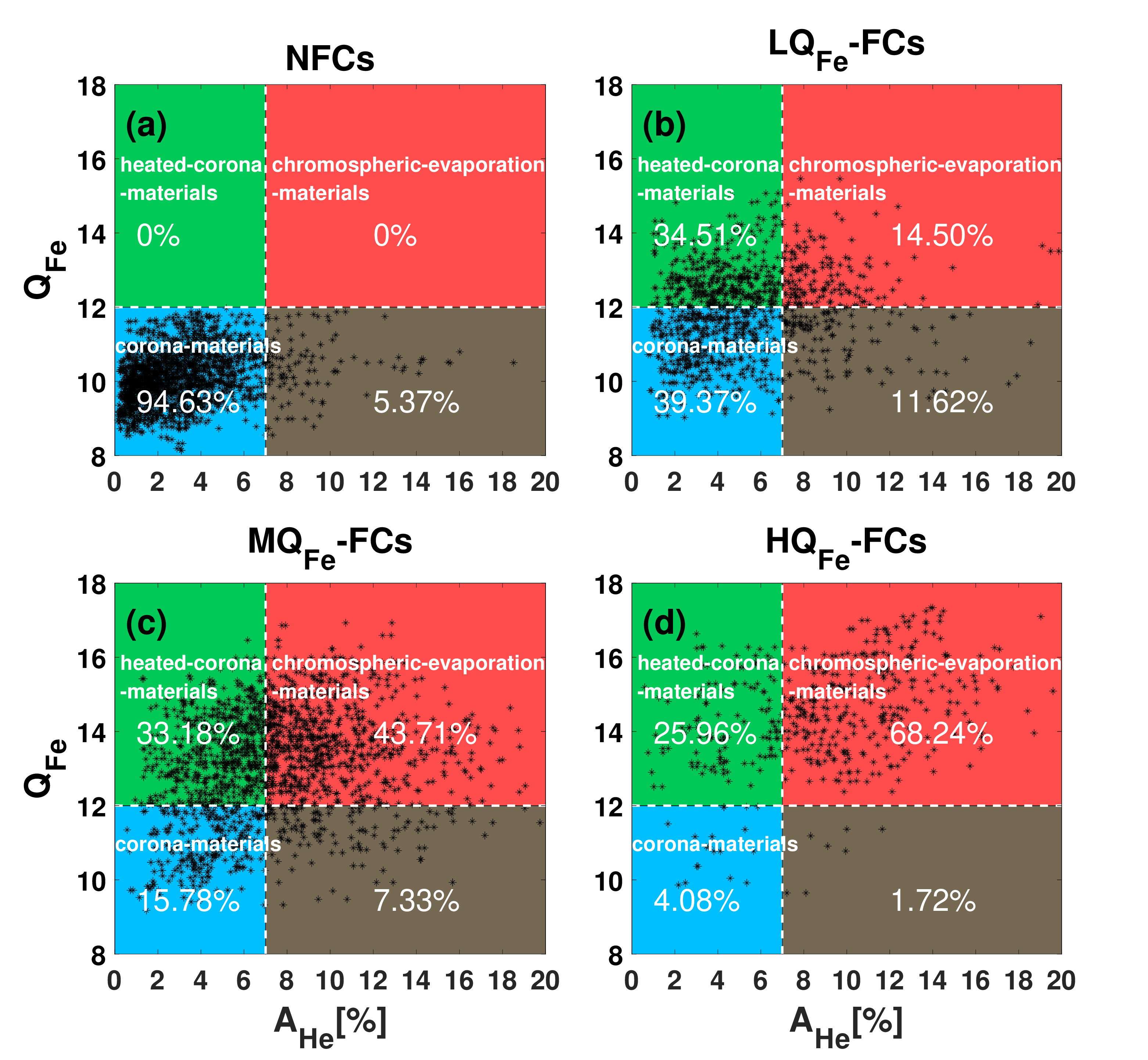}
\caption{The contributions of the three types of materials inside NFCs (panel (a)), LQFe-FCs (panel (b)), MQFe-FCs (panel (c)), and HQFe-FCs (panel (d)). The black dots represent the measures of ICMEs, and the horizontal and vertical white dashed lines denote the threshold for inferring the origination and generation mechanisms of the ICME materials. The subregions with sky blue, green, and red denote the corona-materials (with lower \qfe\ and \ahe), heated-corona-materials (with higher \qfe\ and lower \ahe), and chromospheric-evaporation-materials (with higher \qfe\ and \ahe), respectively. It is clear that the proportions of the three types of materials are significantly different for the ICMEs associated with different flare intensities.}
\label{fig:figure2_contributions}
\end{figure}

The distribution characteristics in \ahe\ and \qfe\ space for NFCs and three types of FCs are presented in Figure~\ref{fig:figure2_contributions}. The \ahe\ and \qfe\ space is divided into four quadrants by the two dashed lines. The horizontal (\qfe=12) and vertical dashed lines (\ahe=7) denote the thresholds for inferring the origination of the materials inside ICMEs \citep{2022ApJ...928..136Z}.

The distributions of NFCs (Figure~\ref{fig:figure2_contributions}(a)) and LQFe-FCs (Figure~\ref{fig:figure2_contributions}(b)), MQFe-FCs (Figure~\ref{fig:figure2_contributions}(c)), and HQFe-FCs (Figure~\ref{fig:figure2_contributions}(d)) are significantly different.
The NFCs are mainly (94.6\%) located in the third quadrant with lower \qfe\ and \ahe. In contrast, the distributions for FCs are more spreading. The proportions of LQFe-FCs (MQFe-FCs) in the first to the {fourth} quadrant are 14.5\% (43.7\%), 34.5\% (33.2\%), 39.4\% (15.8\%), and {11.6\% (7.3\%)}, respectively.
The samples of the HQFe-FCs are concentrated in the first quadrant. The proportions of the materials in the first to {fourth} quadrants are 68.2\%, 26.0\%, 4.1\%, and {1.7\%}, respectively.

The materials in the first to the third quadrant should come from different regions and be heated by different mechanisms \citep{2020ApJ...900L..18F, 2022ApJ...928..136Z}.
The materials in the third quadrant (with lower \qfe\ and \ahe) should come from the corona directly. The CMEs are generally associated with coronal dimming, which is caused by the loss of coronal materials \citep{1996ApJ...470..629H, 2000A&A...358.1097H, 2003A&A...400.1071H, 2018ApJ...863..169D}.
If the materials are not heated during the flaring processes, their \qfe\ should be lower. The \ahe\ should also be lower as the helium abundance is depleted in the corona \citep{2001ApJ...546..552L, 2003ApJ...591.1257L}.
The materials in the second quadrant (with higher \qfe\ and lower \ahe) should be heated by magnetic reconnection directly in the corona during the flaring processes.
The materials can be heated by the magnetic reconnection taking place between the outward flux rope and flare loops during the flaring processes \citep{2000JGR...105.2375L, 2013NatPh...9..489S, 2015SSRv..194..237L}. In this scenario, the \qfe\ is higher as the materials are strongly heated, and the \ahe\ should be lower as the materials are located at the corona.
The materials in the first quadrant (with higher \qfe\ and \ahe) should be heated by the chromospheric evaporation process at the flux rope footpoint regions \citep{2020ApJ...900L..18F, 2022ApJ...928..136Z}.
\citet{2020ApJ...900L..18F} confirm that the plasma heated by the chromospheric evaporation processes can also be transported into CMEs. The higher \ahe\ inside ICMEs can be explained reasonably by the above mass supply scenario, as the \ahe\ in the lower solar atmosphere should be higher than that in the corona.
{The statistical results based on only 10 flare-ICMEs detected from 2009 to 2013 demonstrate that the contribution of the materials generated by the chromospheric evaporation process can not be ignored \citep{2022ApJ...928..136Z}.}
{Further research is needed to investigate the origination and generation mechanisms of the plasma in the fourth quadrant.
The plasma in the fourth quadrant is associated with lower \qfe\ and higher \ahe.
The materials in the fourth quadrant are not strongly heated on the Sun as the \qfe\ is lower.
The materials may come from the lower solar atmosphere without strong heating.
On the other hand, the plasma may be closely related to the filament, considering that the \ahe\ in the lower solar atmosphere and filament should be higher than that in the corona.
However, the \qfe\ for the plasma in the fourth quadrant is nearly the same as that in the third quadrant.
This is not consistent with the notion that the charge states of the materials related to the filament should be extremely low \citep{2010ApJ...723L..22L}.}

The present statistical results demonstrate that the material origination and generation mechanisms of ICMEs are significantly influenced by associated flare intensities.
The materials inside NFCs are mainly concentrated in the third quadrant with lower \qfe\ and \ahe\ (Figure~\ref{fig:figure2_contributions}(a)). This is consistent with the notion that the materials of ICMEs not associated with flares should come from the corona directly \citep{1996ApJ...470..629H, 2000A&A...358.1097H, 2003A&A...400.1071H, 2018ApJ...863..169D, 2022ApJ...928..136Z}.
In contrast, the material sources and generation mechanisms are more diverse for the ICMEs associated with flares. The contributions of all three types of materials inside FCs cannot be ignored. Quanlititively, the contribution of materials coming from the corona directly decreases with flare intensities. The proportions of heated plasma (with higher \qfe) inside ICMEs significantly increase with flare intensities. The proportions of heated materials (associated with higher \qfe) for LQFe-FCs, MQFe-FCs, and HQFe-FCs are 49.0\%, 76.9\%, and 94.2\%, respectively.
{The statistical results demonstrate that the stronger the flares, the more heated materials inside ICMEs.}

The proportions of two types of heated materials can be quantitively elevated in the present study (see Panels (b)-(d) in Figure~\ref{fig:figure2_contributions}). About one-third of materials inside FCs are heated by magnetic reconnection in the corona, and its proportions are slightly decreased with the increase of flare intensities. In contrast, the contribution of hot materials produced by chromospheric evaporation processes is significantly increased with flare intensities.
The proportions of heated-corona-materials (chromospheric-evaporation-materials) for LQFe-FCs, MQFe-FCs, and HQFe-FCs are 34.5\% (14.5\%), 33.2\% (43.7\%), and 26.0\% (68.2\%), respectively.
The present statistical results demonstrate that the two material-heated mechanisms of ICMEs are both important.
{More than two-thirds of hot materials are heated by magnetic reconnection directly in the corona for ICMEs associated with weaker flares.
In contrast, the majority of materials inside ICMEs associated with strong flares are generated by the chromospheric evaporation processes at flux rope footpoint regions.}

\subsection{The FIP bias of the three types of materials inside ICMEs that are associated with different flare intensities}

\begin{figure}
\centering
\includegraphics[width=1.0\textwidth]{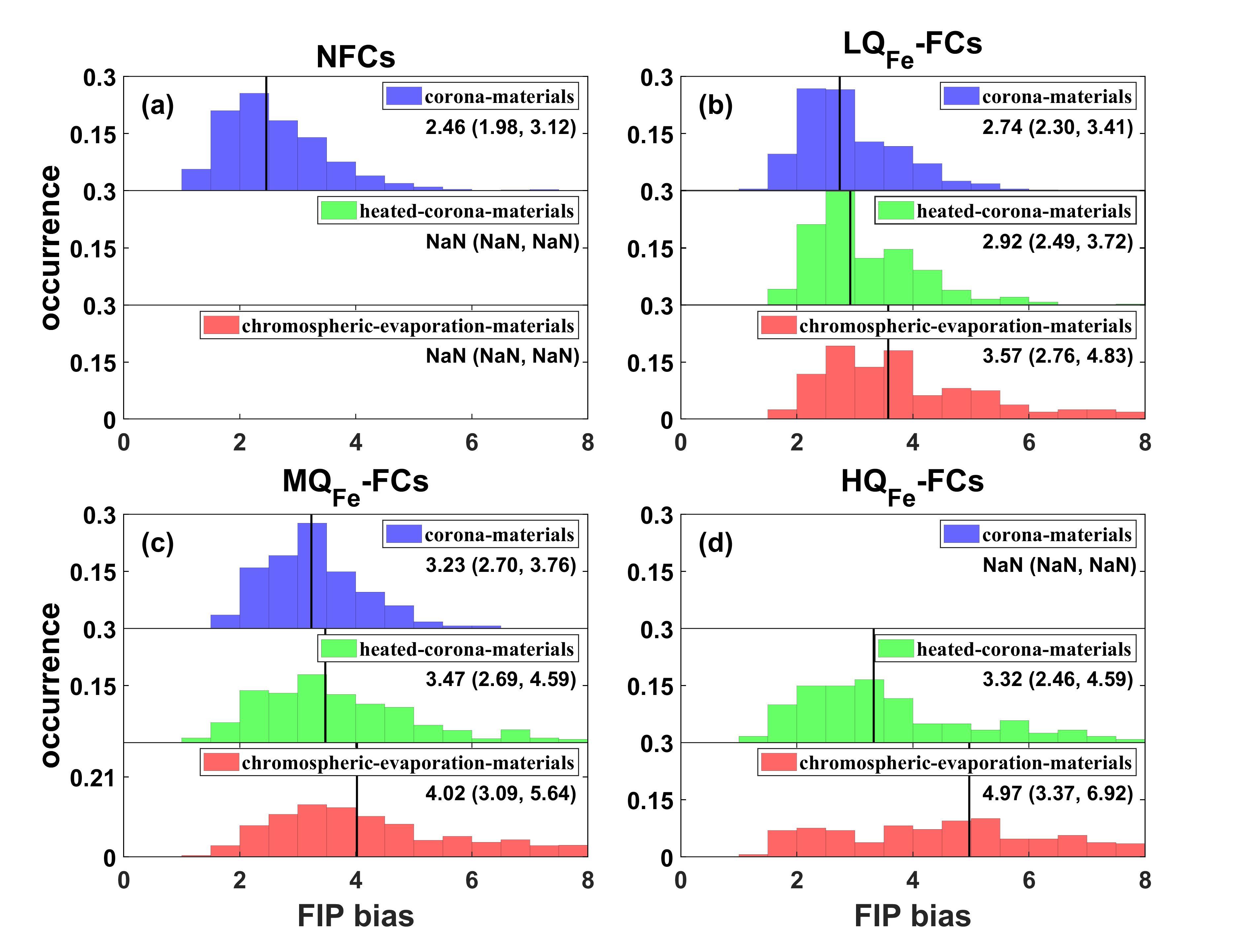}
\caption{The FIP bias of the three types of materials inside NFCs (panel (a)), LQFe-FCs (panel (b)), MQFe-FCs (panel (c)), and HQFe-FCs (panel (d)).
{The distributions are normalized by the samples of the present type of materials, as we concentrate on the comparison of the distribution characteristics between different types of ICME materials.}
The blue, green, and red bars represent the corona-materials, heated-corona-materials, and chromospheric-evaporation-materials, respectively. The black vertical lines in each panel denote the {median} FIP bias. {The median and lower and upper quartiles of FIP bias are also presented in each panel.} We can see that the FIP bias of chromospheric-evaporation-materials is higher than that of corona-materials and heated-corona-materials in all three types of FCs.}
\label{fig:figure3_FIP_bias}
\end{figure}

In this section, we concentrate on the FIP bias of the three types of materials (corona-materials, heated-corona-materials, and chromospheric-evaporation-materials) inside ICMEs associated with different flare intensities.
The FIP bias of NFCs, LQFe-FCs, MQFe-FCs, and HQFe-FCs is given in Figure~\ref{fig:figure3_FIP_bias}. In each panel, the histograms of the FIP bias for the three types of materials that come from different sources and/or are generated by different mechanisms are presented.

There are three distinct characteristics of the FIP bias inside three types of materials.
First, the FIP bias of corona-materials (blue bars in Figure~\ref{fig:figure3_FIP_bias}) is almost the same as that of heated-corona-materials (green bars in Figure~\ref{fig:figure3_FIP_bias}).
Second, the FIP bias of the chromospheric-evaporation-materials (red bars in Figure~\ref{fig:figure3_FIP_bias}) is higher than that of corona-materials and heated-corona-materials. The above characteristics can be clearly seen in panels (b)-(d) in Figure~\ref{fig:figure3_FIP_bias}, in which the histograms of corona-materials and heated-corona-materials are nearly the same, and the distributions of chromospheric-evaporation-materials in all three types of FCs all lie at the right.
Third, the FIP bias of chromospheric-evaporation-materials is increased with the increase of flare intensities.
{The median and lower and upper quartiles of FIP bias for corona-materials inside NFCs are 2.46 (1.98, 3.12), and the median and lower and upper quartiles of FIP bias for corona-materials, heated-corona-materials, and chromospheric-evaporation-materials inside LQFe-FCs are 2.74 (2.30, 3.41), 2.92 (2.49, 3.72) and 3.57 (2.76, 4.83), respectively.
The median and lower and upper quartiles of FIP bias for three types of materials inside MQFe-FCs are 3.23 (2.70, 3.76), 3.47 (2.69, 4.59), and 4.02 (3.09, 5.64).
The median and lower and upper quartiles of FIP bias of heated-corona-meterials and chromospheric-evaporation-materials inside HQFe-FCs are 3.32 (2.46, 4.59) and 4.97 (3.37, 6.92).}

The above characteristics of FIP bias
can be explained reasonably by the origination and generation mechanisms of three types of ICME materials.
The remote spectroscopy observations demonstrate that the FIP bias is not uniformly distributed in the corona \citep{2001ApJ...555..426W, 2005JGRA..110.7109F, 2013ApJ...778...69B, 2015NatCo...6.5947B}. Generally, the FIP bias in active regions is higher than that of quiet Sun and coronal hole regions \citep{2003SSRv..107..665F, 2005JGRA..110.7109F, 2006ApJ...646.1275K}, although the FIP bias changes with the lifetime of the corona structures \citep{2001ApJ...555..426W, 2003SSRv..107..665F, 2015ApJ...802..104B, 2018ApJ...856...71B}.

The FIP bias statistical results of corona-materials, heated-corona-materials, and chromospheric-evaporation-materials inside ICMEs are consistent with the remote measures of the FIP bias on the source regions.

First, the FIP bias inside corona-materials and heated-corona-materials is almost the same as they both come from the corona. The essential difference between corona-materials and heated-corona-materials is that the latter (former) is (not) heated by magnetic reconnection during flaring processes. Qualitatively, the FIP bias should not be changed by the above sudden heating processes. Therefore, the FIP bias inside corona-materials and heated-corona-materials is nearly the same.

Second, the FIP bias of chromospheric-evaporation-materials is the highest as they originate from the flux rope footpoint regions. The spectroscopy observations show that the FIP bias is the highest in the active region loop footpoint regions, which correspond to the magnetic concentrated areas (see Figures~2 and 3 in \citeauthor{2013ApJ...778...69B}, \citeyear{2013ApJ...778...69B} and Figure~2 in \citeauthor{2015ApJ...802..104B}, \citeyear{2015ApJ...802..104B}). The chromospheric-evaporation-materials are generated by chromospheric evaporation processes at the flux rope footpoint regions. Therefore, the FIP bias of chromospheric-evaporation-materials is the highest as they come from loop footpoint regions where the FIP bias is the highest in the corona.

Third, the fact that the FIP bias of chromospheric-evaporation-materials is increased with flare intensities can also be explained reasonably by the origination of chromospheric-evaporation-materials. The chromospheric-evaporation-materials are generated by chromospheric evaporations at flux rope footpoint regions during flaring processes. Qualitatively, the higher the flare intensities, the stronger the active regions and magnetic field concentrated areas.
The remoted observations demonstrate that the FIP bias is the highest at the strong magnetic field concentrated areas \citep{2013ApJ...778...69B, 2015ApJ...802..104B}. Therefore, the FIP bias of chromospheric-evaporation-materials is increased with the increase of the flare intensities, statistically.

The present work advanced our understanding on the complicated properties of ICMEs and on the relationship between CMEs and flares.
The complicated charge states and element abundance inside ICMEs are induced by the diversity and variation of the origination and generation mechanisms of ICME materials. Part of the ICME materials come from the corona with or without strong heating, and the other materials may be heated by the chromospheric evaporation process at the flux rope footpoint regions. The properties and proportions of the above different types of materials inside ICMEs are not the same. Therefore, the characteristics of charge states and element abundance (helium abundance and FIP bias) inside ICMEs are complicated. The in-situ properties of ICMEs are generally different case by case \citep{1997GMS....99..245N, 2003JGRA..108.1156C, 2004JGRA..109.9104R, 2010SoPh..264..189R, 2017SSRv..212.1159M}.

The statistical results also deepen our understanding on the characteristic of FIP bias inside ICMEs. The reason for the elevation of FIP bias inside ICMEs is that the materials generated by chromospheric evaporation processes at flux rope footpoint regions are statistically associated with higher FIP bias.
The previous studies find that the FIP bias inside ICMEs is higher than that of the solar wind in a statistical manner \citep{2004JGRA..109.9104R, 2017SSRv..212.1159M, 2018SoPh..293..122O, 2021SoPh..296..111S, 2022ApJ...940..103S}. In addition, the FIP bias of the ICMEs associated with flares is significantly higher than that of ICMEs without flares \citep{2022ApJ...940..103S}. The reason is that the materials generated by chromospheric evaporation processes at flux rope footpoint regions are associated with higher FIP bias, and the contribution of the above higher FIP bias materials is significantly increased with flare intensities.

\section{Summary and Conclusions}

In the present study, the ICMEs associated with different flare intensities are analyzed and compared. The materials inside ICMEs are first classified into three types, coming from the corona directly (corona-materials), heated by magnetic reconnection in the corona (heated-corona-materials), and generated by chromospheric evaporation processes in the chromosphere (chromospheric-evaporation-materials) based on in-situ detected \qfe\ and \ahe\ \citep{2020ApJ...900L..18F, 2022ApJ...928..136Z}. Then the origination and properties of three types of materials, which come from different regions and/or are generated by different mechanisms, are statistically analyzed and compared for the ICMEs associated with different flare intensities. The main results are concluded as follows:

\begin{enumerate}
\item
The speeds and scales of near-Earth ICMEs both increase with flare intensities. The average speeds (scales) of HQFe-FCs are significantly higher (larger) than those of NFCs, LQFe-FCs, and MQFe-FCs. The speeds (scales) of NFCs, LQFe-FCs, MQFe-FCs, and HQFe-FCs are 424$\pm$101 \velunit\ (0.24$\pm$0.12 AU), 476$\pm$107 \velunit\ (0.27$\pm$0.13 AU), 498$\pm$103 \velunit\ (0.38$\pm$0.21 AU), and 601$\pm$120 \velunit\ (0.47$\pm$0.30), respectively.

\item
The proportions of the three types of materials are significantly different for the ICMEs associated with different flare intensities.
The proportions of heated-corona-materials are nearly constant with flare intensities. In contrast, the contributions of corona-materials (chromospheric-evaporation-materials) are significantly decreased (increased) with flare intensities. More than two-thirds of materials are generated by chromospheric evaporation processes for the ICMEs associated with stronger flares. The proportions of corona-materials (heated-corona-materials, chromospheric-evaporation-materials) for NFCs, LQFe-FCs, MQFe-FCs, and HQFe-FCs are 94.6\% (0.0\%, 0.0\%), 39.4\% (34.5\%, 14.5\%), 15.8\% (33.2\%, 43.7\%), and 4.1\% (26.0\%, 68.2\%), respectively.

\item
The FIP bias of the three types of materials inside ICMEs is different. The FIP bias of corona-materials and heated-corona-materials is almost the same. Whereas, the FIP bias of chromospheric-evaporation-materials is higher than that of corona-materials and heated-corona-materials, and it is clearly increased with flare intensities. The {median} FIP bias of corona-materials (heated-corona-materials, chromospheric-evaporation-materials) for NFCs, LQFe-FCs, MQFe-FCs, and HQFe-FCs is {2.46 (NaN, NaN), 2.74 (2.92, 3.57), 3.23 (3.47, 4.02), and NaN, (3.32, 4.97), respectively.}
\end{enumerate}

Our results demonstrate that the properties of ICMEs are closely related to the associated flare intensities. The speeds and scales of near-Earth ICMEs are both increased with flare intensities.
The origination and generation mechanisms of ICME materials are also significantly influenced by associated flare intensities. The materials of NFCs mainly come from the corona directly. Quanlititively, the contribution of the materials coming from the corona directly (heated during flaring processes) significantly decreases (increases) with flare intensities.
{In addition, the proportions of the materials heated by magnetic reconnection in the corona are slightly decreased with the increase of flare intensities.
The contribution of the materials produced by chromospheric evaporation processes at flux rope footpoint regions is significantly increased with flare intensities.}
The majority of the materials are generated by chromospheric evaporation processes for the ICMEs associated with strong (X-class) flares.

The characteristics of FIP bias inside three types of materials can be explained reasonably by the origination and generation mechanisms of three types of ICME materials.
The reason for the elevation of FIP bias inside ICMEs is that the materials generated by chromospheric evaporation processes at flux rope footpoint regions are associated with higher FIP bias. In addition, the contribution of the higher FIP bias materials generated by chromospheric evaporation processes is significantly increased with flare intensities. Hence, the FIP bias inside ICMEs is higher than the solar wind, and it is increased with flare intensities if the ICMEs are regarded as a whole.
The present statistical results indicate that the complicated charge states and element abundance inside ICMEs should be induced by the diversity and variation of origination and generation mechanisms of ICME materials.

\begin{acknowledgments}
{The authors thank very much the anonymous referee for the helpful and constructive comments and suggestions.}
We thank the ACE SWICS, 
and SWEPAM instrument teams and the ACE Science Center for providing the ACE data. 
Analysis of Wind SWE observations is supported by NASA grant NNX09AU35G.
This research is supported by the National Natural Science Foundation of China (42230203, 41974201).
\end{acknowledgments}

\bibliography{export-bibtex}

\bibliographystyle{aasjournal}
\end{document}